\documentclass[aip,jcp,reprint,amsmath,amssymb]{revtex4-2}

\usepackage{graphicx}
\usepackage{dcolumn}
\usepackage{bm}
\usepackage[mathlines]{lineno}

\usepackage[utf8]{inputenc}
\usepackage[T1]{fontenc}
\usepackage{mathptmx}
\usepackage{etoolbox}

\makeatletter
\def\@email#1#2{%
 \endgroup
 \patchcmd{\titleblock@produce}
  {\frontmatter@RRAPformat}
  {\frontmatter@RRAPformat{\produce@RRAP{*#1\href{mailto:#2}{#2}}}\frontmatter@RRAPformat}
  {}{}
}%
\makeatother
\begin{document}

\title[]{Nanoscale Protein Diffusion in Supercooled Cryoprotectant Solutions}
\author{Maddalena Bin}
\affiliation{Department of Physics, AlbaNova University Center, Stockholm University, S-106 91 Stockholm, Sweden}
\author{Anita Girelli}
\affiliation{Department of Physics, AlbaNova University Center, Stockholm University, S-106 91 Stockholm, Sweden}
\affiliation{Institut für Angewandte Physik, Universität Tübingen, Auf der Morgenstelle 10, 72076 Tübingen, Germany}
\author{ Mariia Filianina}
\author{ Mario Reiser}
\author{ Sharon Berkowicz}
\author{ Milla Åhlfeldt}
\affiliation{Department of Physics, AlbaNova University Center, Stockholm University, S-106 91 Stockholm, Sweden}

\author{ Michelle Dargasz}
\author{ Sonja Timmermann}
\affiliation{Department Physik, Universität Siegen, Walter-Flex-Strasse 3, 57072, Siegen, Germany}
\author{ Jaqueline Savelkouls}
\affiliation{Fakultät Physik/DELTA, TU Dortmund, Maria-Goeppert-Mayer-Str. 2, 44221 Dortmund, Germany}

\author{ Takeshi Kawasaki}
\affiliation{D3 Center, The University of Osaka, Toyonaka, Osaka, 560-0043, Japan}
\affiliation{Department of Physics, The University of Osaka, Toyonaka, Osaka, 560-0043, Japan}

\author{ Shinji Saito}
\affiliation{Institute for Molecular Science, Myodaiji, Okazaki, Aichi 444-8585, Japan}
\affiliation{The Graduate University for Advanced Studies (SOKENDAI), Myodaiji, Okazaki, Aichi 444-8585, Japan}

\author{ Federico Zontone}
\author{ Yuriy Chushkin}
\affiliation{ESRF - The European Synchrotron, 71 Avenue des Martyrs, Grenoble, 38000, France}

\author{ Fajun Zhang}
\author{ Frank Schreiber}
\affiliation{Institut für Angewandte Physik, Universität Tübingen, Auf der Morgenstelle 10, 72076 Tübingen, Germany}
\author{ Michael Paulus}
\affiliation{Fakultät Physik/DELTA, TU Dortmund, 44221 Dortmund, Germany}
\author{ Christian Gutt}
\affiliation{Department Physik, Universität Siegen, Walter-Flex-Strasse 3, 57072, Siegen, Germany}
\author{ Fivos Perakis}
\affiliation{Department of Physics, AlbaNova University Center, Stockholm University, S-106 91 Stockholm, Sweden}
\email{f.perakis@fysik.su.se}

\begin{abstract}
Vitrification during cryopreservation requires a quantitative understanding of protein transport in deeply supercooled cryoprotectant solutions, yet direct measurements at molecular length scales remain scarce. Here, we combine X-ray Photon Correlation Spectroscopy (XPCS) and small-angle X-ray scattering (SAXS) to investigate ferritin diffusion in glycerol–water mixtures from ambient conditions down to $210$~K. The measured diffusion coefficients reveal that ferritin retains a higher mobility upon cooling than expected from hydrodynamic scaling based on measurements of larger silica reference tracers, with the difference emerging below approximately $230$~K. A minimal fluctuating-friction model reproduces the observed relative enhancement in diffusion, illustrating how local variations in the effective friction can give rise to such behavior. These measurements provide direct experimental benchmarks for future theoretical and simulation studies aimed at understanding molecular transport in deeply supercooled liquids approaching the glass transition.
\end{abstract}

\maketitle
Understanding protein dynamics in cryoprotected aqueous environments is central to improving cryopreservation strategies for biological and pharmaceutical systems~\cite{murray_chemical_2022}. 
Protein diffusion at low temperatures provides a sensitive probe of kinetically trapped and metastable states, revealing conditions that lead to denaturation or aggregation and thus determine the preservation of biological function during vitrification. 

Glycerol, one of the most effective cryoprotectants, strongly influences both protein dynamics~\cite{caliskan_protein_2003,dirama_coupling_2005} and structure~\cite{hirai_direct_2018,filianina_nanocrystallites_2023}. 
By disrupting the hydrogen-bond network of water, glycerol suppresses ice formation and extends the accessible supercooled regime~\cite{berkowicz_supercritical_2024}. Yet, despite its widespread use, the microscopic mechanisms governing molecular transport in deeply supercooled cryoprotectant solutions remain poorly understood~\cite{zaragoza_anomalous_2024}, particularly in relation to the dramatic slowing down of dynamics approaching the glass transition~\cite{Tanaka2012,Tanaka2019}.


In simple liquids, the translational diffusion coefficient $D(T)$ is often related to the solvent viscosity $\eta$ by the Stokes--Einstein (SE) equation,
\begin{equation}
    D(T) = \frac{k_\mathrm{B} T}{6\pi \eta R_{\rm h}},
\end{equation}
where $R_{\rm h}$ is the hydrodynamic radius and $k_\mathrm{B}$ the Boltzmann constant. 

However, a growing body of work has shown that this relation breaks down in supercooled and glass-forming liquids, reflecting the increasingly complex relationship between molecular transport and structural relaxation as the glass transition is approached.\cite{Tanaka2012,Tanaka2019,charbonneau_hopping_2014,sengupta_breakdown_2013,das_crossover_2022,hodgdon_stokes-einstein_1993,mallamace_considerations_2016}, including supercooled water~\cite{kawasaki_identifying_2017} and protein solutions~\cite{lamanna_solvent_1994,chirico_fractional_1999,joshi_glycerol-slaved_2024}. Understanding how molecular transport evolves upon supercooling remains a central problem in liquid-state physics, closely linked to the dramatic slowing down of dynamics approaching the glass transition.

These deviations have been discussed in terms of several microscopic mechanisms, including dynamical heterogeneity, activated hopping processes, and local structural ordering in supercooled liquids.\cite{ediger_spatially_2000,Tanaka2012,Tanaka2019}, where regions of high and low mobility coexist and evolve on nanoscopic length- and time-scales~\cite{becker_fractional_2006}. 
Probing this coupling between protein motion and solvent heterogeneity directly remains an open challenge.

To address this challenge, we employ X-ray Photon Correlation Spectroscopy (XPCS)~\cite{madsen_structural_2016,moller_x-ray_2019,perakis_towards_2020}, which uses coherent X-rays to probe nanoscale dynamics in complex liquids and biological systems~\cite{reiser_resolving_2022,girelli_microscopic_2021,begam_kinetics_2021,timmermann_x-ray_2023,chushkin_probing_2022,ragulskaya_interplay_2021,anthuparambil_exploring_2023, girelliCoherentXraysReveal2025}. XPCS has emerged as a powerful technique to probe dynamics in soft and condensed matter systems across a wide range of length and time scales \cite{grubel_correlation_2004, madsen_beyond_2010, shpyrko_x-ray_2014, sandy_hard_2018}. 
By resolving temporal correlations of the scattered X-ray intensity, XPCS directly measures collective diffusion on molecular length scales, corresponding to the momentum transfer region near the structural peak. 
This capability makes XPCS ideally suited to study protein mobility in supercooled, cryoprotected solutions, where optical methods such as dynamic light scattering (DLS) are often limited by multiple scattering and opacity. Shorter time- and length-scales are typically addressed with complementary approaches using energy-resolved neutron scattering~\cite{grimaldo_dynamics_2019}.

Here, we investigate the collective diffusion of ferritin proteins in glycerol--water mixtures from ambient to cryogenic temperatures using XPCS, complemented by small-angle X-ray scattering (SAXS) to monitor structural stability. 
Ferritin is a protein cage consisting of a rigid, hollow assembly of subunits with an iron-rich core, and therefore represents a well-defined nanoscale object rather than a typical flexible protein in solution. It serves as an ideal model due to the high X-ray contrast from its iron core, and relevance to biomedical applications such as vaccine design~\cite{rodrigues_functionalizing_2021} and drug delivery~\cite{zhu_ferritin-based_2021}. By accessing momentum transfers near the structure-factor peak, we directly probe the molecular-scale dynamics that govern ferritin motion in supercooled cryoprotectant environments.

\section{Methods}
\noindent
Ferritin solutions were prepared from equine spleen ferritin (Sigma--Aldrich, F4503) at an initial protein concentration of $c = 71$~mg\,mL$^{-1}$ in 150~mM NaCl. 
The stock solution was concentrated by centrifugation for 1~h at 10,000~$g$ using 10~kDa Millipore filters, yielding a ferritin concentration of approximately $c = 730$~mg\,mL$^{-1}$. The molecular mass of ferritin (including the iron core) was taken as $\sim$700 kDa, and uncertainties in concentration were propagated consistently in the volume fraction estimates. 
This concentrated solution was then diluted with glycerol and water to obtain intermediate protein concentrations of $c = 70 \pm 40$~mg\,mL$^{-1}$ (used for the SAXS measurements) and $c = 100 \pm 40$~mg\,mL$^{-1}$ (used for XPCS measurements) in glycerol--water mixtures with a 23~mol\% glycerol fraction. 
At these conditions, the protein volume fraction was $\phi = 0.05$, and the hydrodynamic radius $R_{\rm h} = 7.3$~nm (measured by DLS, see Supplementary Material). 
For comparison, silica nanoparticle suspensions were prepared in the same glycerol--water mixture (23~mol\% glycerol) at a particle volume fraction of $\phi \approx 0.002$. These silica nanoparticles were purchased by Alpha Nanotech with a radius of 50 nm (determined by TEM from the manufacturer and confirmed by DLS), initial concentration 10~mg\,mL$^{-1}$ in Milli-Q water and are hydroxyl-terminated. The particles were further diluted to $\phi \approx 0.002$ to obtain sufficient scattering intensity in the dilute regime. Ferritin was measured at higher volume fraction $\phi \approx 0.05$, which is also assumed to be in the dilute regime, in order to ensure that we have sufficient signal. The comparison between ferritin and silica nanoparticles suggests that nanoscale objects exhibit stronger apparent deviations from SE behavior than larger particles under otherwise similar solvent conditions. We note, however, that the two systems differ not only in particle size, but also in concentration, morphology, and interaction potential. Therefore, the present comparison should be viewed as illustrating a general trend rather than a strictly size-controlled experiment.

\begin{table}[htb]
    \centering
    \begin{tabular}{lcc}
    \hline
    Parameter   & ESRF--ID10  & DELTA--BL2  \\
    \hline
    Photon energy (keV) & 9 & 12 \\
    Beam size & $25\times25~\mathrm{\mu m}^2$ & $0.6\times0.6~\mathrm{mm}^2$ \\
    Flux (ph/s) & $7\times10^{10}$ & $1.7\times10^{9}$ \\
    Detector  & Eiger~500k & MAR345 \\
    Sample--detector distance (m) & 7.11 & 1.659 \\
    \hline
    \end{tabular}
    \caption{Experimental parameters for XPCS (ESRF--ID10) and SAXS (DELTA--BL2) measurements.}
    \label{tab:exp_param}
\end{table}

\noindent
SAXS experiments were performed at beamline BL2 of the Dortmund Synchrotron Radiation Source DELTA~\cite{dargasz_x-ray_2022} using 12~keV photons and a sample--detector distance of 1.659~m. 
XPCS measurements were carried out at the ID10 beamline of the European Synchrotron Radiation Facility (ESRF, proposal~SC-5375), using coherent 9~keV X-rays in SAXS geometry with an Eiger~500k detector positioned 7.11~m from the sample. 
This configuration provided access to a momentum-transfer range of $q = 0.10$--$0.43$~nm$^{-1}$. 
The incident beam had a full-width at half-maximum of $25\times25~\mathrm{\mu m}^2$ and a flux of $7\times10^{10}$~ph\,s$^{-1}$. 
Samples were loaded into 1.5~mm diameter quartz capillaries and mounted in a nitrogen gas cryostat~\cite{steinmann_small-angle_2011}, allowing temperature control down to $T = 193$~K. 
To minimize beam-induced effects, the sample position was refreshed for each exposure. 
For each temperature, approximately 120 positions were recorded per capillary with 60~$\mathrm{\mu m}$ spacing, ensuring sufficient statistics and signal-to-noise.

\section{Results \& Discussion}

\begin{figure}[b]
    \centering
    \includegraphics[width=0.9\linewidth]{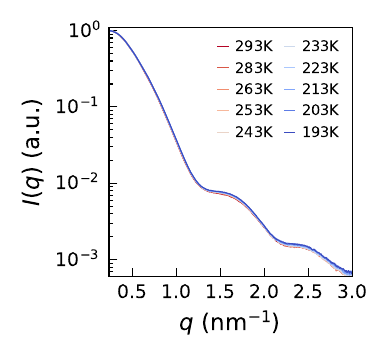}
    \caption{%
    Temperature-dependent SAXS intensity $I(q)$ for ferritin solutions in glycerol--water mixtures (23 mol\% glycerol) and protein concentration $c \approx 70$~mg\,mL$^{-1}$ (volume fraction $\phi \approx 0.05$). The nearly invariant scattering profiles indicate structural stability and the absence of cold denaturation or nanocrystallization upon cooling.%
    }
    \label{fig:saxs}
\end{figure}

\begin{figure*}[tbh]
    \centering
    \includegraphics[width=0.9\linewidth]{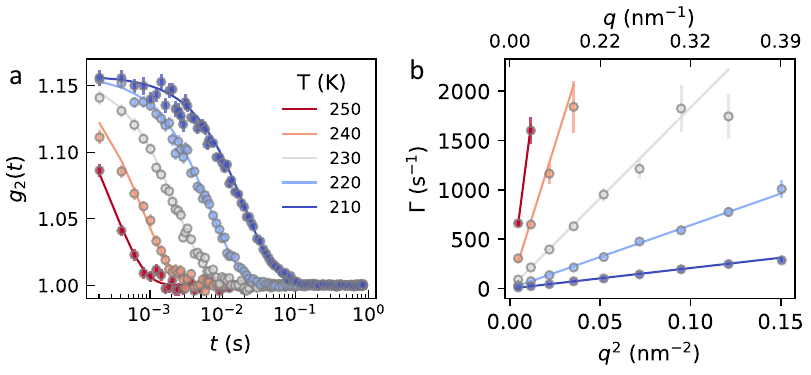}
    \caption{%
    (a)~Intensity autocorrelation functions $g_2(q,t)$ for ferritin solutions at $q = 0.1$~nm$^{-1}$ and various temperatures. 
    Solid lines represent exponential fits. 
    (b)~Relaxation rate $\Gamma(q)$ as a function of $q^2$ for the same temperatures. 
    Linear fits with $\Gamma(q)=Dq^2$ yield the diffusion coefficients $D(T)$. 
    Error bars reflect fitting uncertainties propagated from the exponential model.%
    }
    \label{fig:g2s}
\end{figure*}

Figure~\ref{fig:saxs} shows the small-angle X-ray scattering (SAXS) intensity $I(q)$ of ferritin in glycerol--water mixtures measured between $T = 293$~K and $T = 193$~K. 
The momentum transfer is defined as $q = 4\pi \sin(\theta) / \lambda$, where $\lambda$ is the X-ray wavelength and $2\theta$ the scattering angle. 
The data, recorded for a ferritin concentration of $c \approx 70$~mg\,mL$^{-1}$ (solid lines) and corrected for background scattering, display no significant change in the overall $I(q)$ profile upon cooling. 
The absence of major temperature-dependent changes indicates that the protein--protein interactions remain largely unchanged across the studied temperature range. 
Subtle variations in the intensity, on the order of 1\% of the total signal, are attributed to temperature-dependent changes in the solvent compressibility~\cite{berkowicz_supercritical_2024}. 
No evidence of freezing was observed in the temperature range investigated by XPCS. As shown in Fig.~S5, freezing of ferritin solutions produces pronounced changes in the SAXS profile that are absent in the present data, indicating that the samples remain in the supercooled liquid state throughout the analyzed temperature range, consistent with previous studies~\cite{boutet_precrystallization_2007}.

The nanoscale dynamics of the ferritin solutions were probed using XPCS~\cite{madsen_structural_2016,moller_x-ray_2019,perakis_towards_2020}. 
XPCS measures temporal fluctuations in the scattered X-ray intensity through the normalized intensity autocorrelation function,
\begin{equation}
    g_2(q,t) = \frac{\langle I(q,t_0) I(q,t_0 + t) \rangle}{\langle I(q,t_0) \rangle^2},
\end{equation}
where $I(q,t_0)$ and $I(q,t_0 + t)$ are the intensities recorded at times $t_0$ and $t_0 + t$, respectively, and $\langle \dots \rangle$ denotes an average over time and detector pixels within a $q$-bin. 
The measured $g_2(q,t)$ functions were well described by a single-exponential decay,
\begin{equation}
    g_2(q,t) = 1 + \beta \exp[-2t\,\Gamma(q)],
\end{equation}
where $\Gamma(q)$ is the decorrelation rate and $\beta = 0.16$ is the speckle contrast, determined from static reference measurements. The reported speckle contrast represents the average fitted value over the analyzed $q$-range. The selected $q$-ranges were chosen to maximize the overall quality of the XPCS data, while the speckle contrast varies only weakly within these ranges and shows small differences between experimental datasets.

Figure~\ref{fig:g2s}a shows the intensity autocorrelation functions $g_2(q,t)$ measured at $q = 0.1$~nm$^{-1}$ for temperatures between $T = 250$~K and $T = 210$~K. Due to the strong scattering contrast of the ferritin iron core, the measured XPCS signal is dominated by the particle contribution, and thus primarily reflects the single-particle dynamics of ferritin. The lower temperature limit of the XPCS measurements is set by the onset of freezing under the experimental conditions, which limits the temporal stability required for correlation analysis, whereas SAXS measurements probe instantaneous structure on shorter timescales and can therefore be extended to lower temperatures.
The decay of $g_2(q,t)$ continuously slows upon cooling, reflecting the temperature dependence of the ferritin dynamics. 
The corresponding relaxation rates $\Gamma(q)$, plotted as a function of $q^2$ in Fig.~\ref{fig:g2s}b, follow the relation $\Gamma(q) = D q^2$, confirming purely diffusive behavior across the investigated $q$-range. The speckle contrast exhibits a weak $q$-dependence due to the finite coherence and bandwidth of the X-ray beam; however, for the Si(111) monochromator used here ($\Delta E/E \approx 1.4 \times 10^{-4}$), this variation is expected to be small over the investigated $q$-range ~\cite{perakis_towards_2020}.
Beam-induced effects were excluded through flux-dependent control measurements (see Supplementary Material).

\begin{figure*}[tbh]
    \centering
    \includegraphics[width=0.8\linewidth]{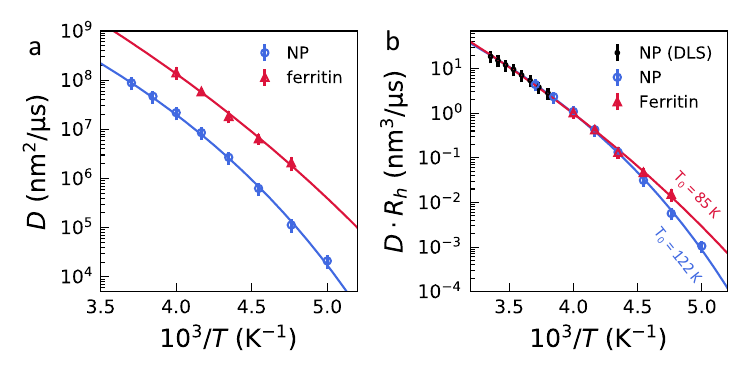}
    \caption{(a) Diffusion coefficient of ferritin solutions (red triangles) compared to nanoparticle solutions (blue circles) obtained from XPCS measurements between $T = 250$ K and $T = 210$ K. Solid lines are fits using the Vogel-Fulcher-Tammann (VFT) relation. %
    (b) The diffusion coefficients  multiplied with the hydrodynamic radius $R_{\rm h}$ (for ferritin $R_{\rm h}=7.3$ nm and for nanoparticles $R_{\rm h}=50$ nm). DLS measurements of the nanoparticle diffusion coefficient are also shown (blue squares). Solid lines are the VFT fits, where the arrest temperature $T_0$ is indicated.   }
    \label{fig:diffusion}
\end{figure*}

The extracted diffusion coefficients $D(T)$ are compared in Fig.~\ref{fig:diffusion}a for ferritin (red triangles) and silica nanoparticle reference solutions (blue symbols).  While ferritin and silica nanoparticles differ in surface chemistry (proteinaceous vs hydroxylated inorganic interfaces), the comparison here is based on their hydrodynamic size and diffusive behavior, providing an experimental reference for diffusion under identical solvent conditions. 
The ferritin data were obtained by XPCS in the momentum-transfer range $q = 0.1-0.43$~nm$^{-1}$, whereas the nanoparticle data were measured at $q = 0.01-0.05$~nm$^{-1}$ and complemented by dynamic light scattering (DLS) measurements (blue squares; see Supplementary Material). The $q$-ranges were selected to optimize scattering intensity and timescales, while probing comparable values of $qR$ for ferritin and silica nanoparticles, ensuring that both systems are measured at similar relative length scales.
To account for size differences, Fig.~\ref{fig:diffusion}b presents the diffusion coefficients multiplied with the respective hydrodynamic radii, $R_{\rm h} = 7.3$~nm for ferritin and $R_{\rm h} = 50$~nm for the nanoparticles. We observe that the two datasets overlap down to $T=230$~K and deviate below this temperature.

The temperature dependence of $D(T)$ is well captured by the Vogel–Fulcher–Tammann (VFT) relation,
\begin{equation}
    D(T) = A \exp\!\left[-\frac{B\,T_0}{T - T_0}\right],
\end{equation}
where $A$, $B$  and $T_0$
are empirical fitting parameters describing the temperature dependence of the diffusion coefficient \cite{chen_stokes-einstein_2006}. 
The VFT model, which describes the non-Arrhenius slowing down of molecular motion in supercooled liquids~\cite{angell_relaxation_1991,turnbull_free-volume_1961,Tanaka2012}, provides a good fit to both datasets. 
As shown in Fig.~\ref{fig:diffusion}, the fits yield $T_0 = 122 \pm 4$~K for the nanoparticle solutions and $T_0 = 85 \pm 12$~K for the ferritin solutions.  The parameter $T_0$ represents an phenomenological descriptor of the effective dynamical arrest temperature within the VFT framework and the extrapolated values of $T_0$ are used only for comparative purposes and should not be interpreted as directly measured arrest temperatures.

Within the VFT framework, the parameter $B$ can be interpreted as an effective activation energy governing the temperature dependence of the dynamics. The parameter $B$ increases from $B = 11 \pm 1$ for the nanoparticles to $B = 26 \pm 8$ for ferritin. The larger value obtained for ferritin indicates a stronger apparent temperature dependence of nanoscale diffusion compared with the larger silica tracer particles. The microscopic origin of this difference remains an open question and is discussed further below.

\begin{figure}[t]
    \centering
    \includegraphics[width=0.8\linewidth]{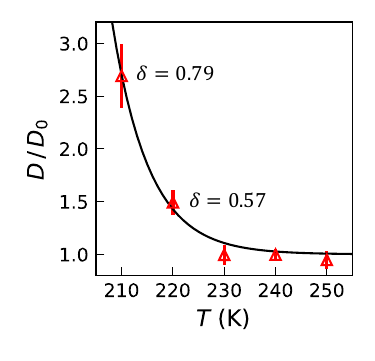}
    \caption{%
    Ratio of the measured ferritin diffusion coefficient $D$ to the reference diffusion coefficient obtained from the silica nanoparticle measurements after scaling by the hydrodynamic radius (defined in Eq.~\ref{eq:ratio}) as a function of temperature $T$.
    The symbols show the experimental data obtained from XPCS measurements, whereas 
    the solid line represents the prediction of the fluctuating--friction model based on Eq.~\ref{eq:fluctuating_friction},
    where $\delta = \Delta\gamma / \gamma_0$ denotes to the relative amplitude of local friction fluctuations.%
    }
    \label{fig:ratio}
\end{figure}

Figure~\ref{fig:ratio} shows the ratio between the measured ferritin diffusion coefficient $D$ and a SE reference $D_0$, defined from the nanoparticle data as
\begin{equation}
D_0 = \frac{R_{\mathrm{h, NP}}}{R_{\mathrm{h,P}}} \, D_{\mathrm{NP}},
\label{eq:ratio}    
\end{equation}

where $R_{\mathrm{h, P}}$ and $R_{\mathrm{h, NP}}$ are the hydrodynamic radii of ferritin and the nanoparticles, respectively, and $D_{\mathrm{NP}}$ is the nanoparticle diffusion coefficient. 
Based on this definition, the $D_0$ corresponds to the diffusion coefficient that ferritin would have if it obeyed the SE relation in the same solvent environment. We assume that the nanoparticles, given the larger size and low concentration, follow the SE relation to a good approximation~\cite{berkowicz_exploring_2021}. 

Pronounced deviations from SE behavior emerge for ferritin below $T \approx 230$~K, where $D/D_0$ increases up to $\sim 2.7$ at $T = 210$~K. We note that the nanoparticle SE relation may itself show minor deviations at the lowest temperatures, but such effects would only reduce the apparent SE violation for ferritin and therefore cannot account for the large enhancement observed. 
Variations in hydrodynamic boundary conditions, from stick to slip~\cite{chirico_fractional_1999}, could alter the SE prefactor from 6 to 4, yet such a change is insufficient to explain the observed enhancement, indicating a genuine breakdown of the SE relation at the molecular scale. A change in hydrodynamic radius sufficient to account for the observed SE deviation would require a reduction by a factor of $\sim$2.7 at the lowest temperatures, which is not consistent with the SAXS data showing no significant structural changes. While small temperature-dependent changes in hydrodynamic radius cannot be excluded, explaining the observed diffusion behavior solely through particle-size changes would require a substantially larger reduction in effective radius. Previous SAXS studies have shown that temperature-dependent hydration effects produce only modest changes in structural parameters, primarily reflecting changes in the hydration shell rather than large changes in particle dimensions\cite{linse_depletion_2025}. Therefore, particle-size variations alone are unlikely to account for the observed SE deviation. Finite-concentration effects are expected to reduce the measured diffusion coefficient relative to the infinite-dilution limit. However, previous MHz-XPCS studies on concentrated ferritin solutions have shown that such concentration-dependent effects remain relatively modest at volume fractions comparable to those investigated here and become pronounced only at substantially higher concentrations, where anomalous diffusion and cage effects emerge~\cite{girelliCoherentXraysReveal2025}. Moreover, these concentration corrections cannot account for the strong temperature dependence of the apparent SE deviation observed in the present work.

The observed relative enhancement of ferritin diffusion indicates that transport on molecular length scales becomes progressively decoupled from that behavior inferred from larger tracer particles under otherwise similar solution conditions, consistent with the broader phenomenology observed in deeply supercooled liquids~\cite{Tanaka2012,Tanaka2019}. Several mechanisms have been proposed to explain such apparent deviations from the SE relation in supercooled liquids, including heterogeneous relaxation, activated hopping processes, and fluctuating local environments. The present measurements do not distinguish uniquely between these mechanisms but provide direct experimental constraints on nanoscale transport in deeply supercooled cryoprotectant solutions.

As one possible interpretation of the observed diffusion enhancement, we consider a minimal fluctuating-friction model in which the local friction experienced by the ferritin is not constant but fluctuates due to temporal and spatial heterogeneities in the surrounding solvent. 
Following the framework of Ref.~\cite{rozenfeld_brownian_1998} the instantaneous friction coefficient $\gamma(t)$ is expressed as 
$\gamma(t)=\gamma_0+\Delta\gamma(t)$, 
where $\gamma_0$ is the average (macroscopic) friction coefficient and $\Delta\gamma(t)$ represents the fluctuations around this mean. This analytical model is consistent with numerical simulations of particle diffusion
in fluctuating environment \cite{perakisEnhancedParticleDiffusion2026}.
The average friction $\gamma_0$ is related to the macroscopic viscosity $\eta$ through the SE 
$\gamma_0 = 6\pi\eta R_{\rm h}$, 
which defines the corresponding SE diffusion coefficient $D_0 = k_{\mathrm{B}}T/\gamma_0$. 
The parameter $\Delta\gamma$ quantifies the local deviations of the instantaneous friction from its mean value and can be expressed in dimensionless form as 
$\delta = \Delta\gamma / \gamma_0$. In the fast-fluctuation limit of $\Delta\gamma(t)$, the protein effectively
experiences the average friction coefficient, and its diffusion coefficient
can be expressed as $D_0$. In contrast, in the slow-fluctuation limit, where the protein cage motion averages over quasi-static domains of different local viscosity, the effective diffusion coefficient is given by (Eq. 43 in Ref.~\cite{rozenfeld_brownian_1998})
\begin{equation}
    \frac{D}{D_0} = \frac{1}{1 - \delta^2}.
    \label{eq:fluctuating_friction}
\end{equation}
This relation links the observed enhancement of $D/D_0$ directly to the magnitude of local friction heterogeneity, 
providing a quantitative framework for relating the measured diffusion enhancement due to fluctuations in the effective local friction.
At $T=220$~K, the extracted value is $\delta \approx 0.57$, while at $T=210$~K it increases to $\delta \approx 0.79$, indicating that the local friction deviates by nearly $\sim80\%$ from the mean friction.

\section{Conclusions}

\noindent By combining XPCS and SAXS, we resolve nanoscale dynamics of ferritins in glycerol--water solutions across the supercooled regime. 
This approach closes a critical gap in understanding protein motion in cryoprotected environments and demonstrates that XPCS can directly access molecular diffusion at cryogenic temperatures. 
The SAXS data confirm that the structural integrity of ferritin is preserved throughout the temperature range, with no indication of cold denaturation or nanocrystallite formation. 

The comparison with larger silica nanoparticles indicates that nanoscale ferritin diffusion retains a higher mobility upon cooling than expected from simple hydrodynamic scaling. The fluctuating-friction model demonstrates that local variations in the effective friction provide one possible framework capable of reproducing the observed enhancement in ferritin diffusion. Although the present measurements do not uniquely identify the microscopic mechanism responsible for the pronounced deviation from SE behavior below $T \approx 230$~K, they provide direct experimental constraints for future theoretical and simulation studies of nanoscale transport in supercooled cryoprotectant solutions. More generally, the present work establishes XPCS as a powerful approach for studying molecular transport under cryogenic conditions.

\section*{Supplementary Material}
The Supplementary Material contains details of the q-dependent analysis, alternative fitting procedures, XPCS intensity autocorrelation functions for the silica nanoparticle solutions, DLS measurements of the hydrodynamic radius, flux-dependent control measurements, and a comparison of SAXS profiles in the liquid and frozen states.

\section*{acknowledgements}
We acknowledge the European Synchrotron Radiation Facility (ESRF) for provision of synchrotron radiation facilities at ID10 beamline (proposal number SC-5275 and SC-5359) and would like to thank the staff for their assistance. We also thank the DELTA machine group for providing synchrotron radiation for sample characterization. FP acknowledges financial support by the Swedish National Research Council (Vetenskapsrådet) under Grant No. 2019-05542, 2023-05339 and within the Röntgen-Ångström Cluster Grant No. 2019-06075, and the kind financial support from Knut och Alice Wallenberg foundation (WAF, Grant. No. 2023.0052). This research is supported by the Center of Molecular Water Science (CMWS) of DESY in an Early Science Project, the MaxWater initiative of the Max-Planck-Gesellschaft, Carl Tryggers (Project No. CTS21:1589) and the Wenner-Gren Foundations (Project No. UPD2021-0144). F.P. and A.G. acknowledge funding from the European Union’s Horizon Europe research and innovation program under the Marie Skłodowska-Curie grant agreement No. 101149230 (CRYSTAL-X). We also acknowledge BMBR ErUM-Pro funding (05K22PS1, CG),  BMBF (05K19PS1 and 05K20PSA, CG; 05K19VTB, FS and FZ), DFG-ANR (SCHR700/28-1, SCHR700/42-1, FS and FZ). CG and FS acknowledge financial support by the consortium DAPHNE4NFDI in association with the German National Research Data Infrastructure (NFDI) e.V. - project number 46024879. SS acknowledges financial support by JSPS through the Grant-in-Aid for Scientific Research (JP21H04676 and JP23K17361). TK acknowledges support by the JST FOREST Program (Grant No. JPMJFR212T), AMED Moonshot Program (Grant No. JP22zf0127009), JSPS KAKENHI (Grant No. JP24H02203), and Takeda Science Foundation. 

\section*{AUTHOR DECLARATIONS}
\subsection*{Conflict of Interest}
The authors have no conflicts to disclose.


\section*{DATA AVAILABILITY}
Raw data were generated at the European Synchrotron Radiation Facility (ESRF) at ID10 beamline (proposal number SC-5275 and SC-5359). Derived data supporting the findings of this study are available from the corresponding author
upon reasonable request.

\bibliography{ref}

\end{document}